\documentclass[%
 reprint,
 showpacs,
 preprintnumbers,
 amsmath,
 amssymb,
 aps,
 pre,
]{revtex4-1}

\usepackage{graphicx}
\usepackage{dcolumn}
\usepackage{bm}
\usepackage{hyperref}

\usepackage{algorithm}
\usepackage{algpseudocode}
\usepackage{todonotes}

\usepackage{silence}
\newcommand{\kB}{k_{\textrm{B}}}

\renewcommand{\v}[1]{\ensuremath{\mathbf{#1}}} 
\let\crossproduct=\times
\renewcommand{\times}{\cdot} 
\renewcommand{\vec}[1]{\v{#1}} 
\newcommand{\gv}[1]{\ensuremath{\mbox{\boldmath$ #1 $}}} 

\newcommand{\pd}[2]{\frac{\partial #1}{\partial #2}}

\newcommand{\pdd}[2]{\frac{\partial^2 #1}{\partial #2^2}}

\newcommand{\grad}[1]{\gv{\nabla} #1} 
\renewcommand{\div}[1]{\gv{\nabla} \cdot #1} 

\newcommand{\laplacian}[1]{\grad^2 #1}

\newcommand{\Vstr}{V_{\textrm{str}}}
\newcommand{\Fcc}{F_{\textrm{cc}}}

\begin{document}

\preprint{APS/123-QED}

\title{Electrohydrodynamic channeling effects in narrow fractures and pores}

\author{Asger Bolet}
\author{Gaute Linga}%
\author{Joachim Mathiesen}
 \email{mathies@nbi.dk}
\affiliation{%
 Niels Bohr Institute, University of Copenhagen, Blegdamsvej 17, DK-2400 Copenhagen, Denmark
}%

\date{\today}

\begin{abstract}
In low-permeability rock, fluid and mineral transport occur in pores and fracture apertures at the scale of micrometers and below.
At this scale, the presence of surface charge, and a resultant electrical double layer, may considerably alter transport properties.
However, due to the inherent non-linearity of the governing equations, numerical and theoretical studies of the coupling between electric double layers and flow have mostly been limited to two-dimensional or axisymmetric geometries.
Here, we present comprehensive three-dimensional simulations of electrohydrodynamic flow in an idealized fracture geometry consisting of a sinusoidally undulated bottom surface and a flat top surface. We investigate the effects of varying the amplitude and the Debye length (relative to the fracture aperture) and quantify their impact on flow channeling.
The results indicate that channeling can be significantly increased in the plane of flow.
Local flow in the narrow regions can be slowed down by up to $5 \%$ compared to the same geometry without charge, for the highest amplitude considered.
This indicates that electrohydrodynamics may have consequences for transport phenomena and surface growth in geophysical systems.
\end{abstract}

\pacs{Valid PACS appear here}
\maketitle


\section{Introduction}
\label{sec:introduction}
Electric double layers (EDL) play an important role in many chemical and physical processes, and is a controlling factor in many industrially applied microfluidic devices \cite{stone2004} and electrochemical cells \cite{van2010}.
Examples include nanofluidic devices for electrophoretic separation or the large-scale harvesting of energy by mixing fluids of different salinity (``Blue energy'') \cite{siria2017}.
In biological systems, EDLs are important e.g.\ for ion transport across membranes or for polymer aggregation \cite{chen1997,cardenas2000,roux2004}.
In fluid-saturated low-permeability rock, the presence of an EDL can significantly alter the mineral transport and thereby inhibit or amplify transformation reactions, as demonstrated by field observations and nanopore molecular dynamics simulations \cite{plumper2017}.
Furthermore, EDLs alter the effective wetting properties of mineral surfaces (see e.g.\ \cite{hassenkam2011} for a study of reservoir sandstone), which could play an important role in enhanced oil recovery based on injection of low salinity fluids. 

The transport of fluid and minerals in fluid saturated porous rock often occurs in networks of narrow fractures or pores, many of which have (sub) micrometer-sized apertures.
When the pore walls are charged, and the resulting EDL extends significantly into the pore fluid, it may change the bulk flow properties of single fractures and pores, and consequently of the whole network.
Electrokinetic flow, however, is a highly non-linear process, which is hard to quantitatively describe in even the most simple systems.
In general, mean-field approximations are often used to model systems beyond the nanometer range \cite{fiorentino2016a,fiorentino2016b}.
From a number of simplifying assumptions, e.g.\ neglecting ion-ion correlations and non-Coulomb forces (so-called Gouy--Chapman theory), one obtains field equations, which can be used for basic theoretical considerations.
Even then, only simple geometries permit analytical solutions, such as cylindrical capillaries \cite{rice1965}.
In equilibrium and when the electric field is weak, the linearized Poisson--Boltzmann equation can be applied:
\begin{equation}
	\laplacian \varphi = \kappa^2 \varphi,
    \label{eq:poissonboltzmann}
\end{equation}
where $\varphi$ is the electric potential and $\kappa^{-1}$ is the Debye length characterizing the extent of the EDL.
However, when ion transport is coupled to fluid advection, the equilibrium assumption generally breaks down and other means must be pursued.
Further, numerical simulations can be challenging, and have in general been limited to simple geometries such as finite-length symmetric channels e.g.\ in studies of transient streaming potentials in single-phase flow \cite{mansouri2005,mansouri2007} or in studies of electroconvection near perm\-selective membranes \cite{demekhin2011,druzgalski13}.

%


Here, we consider electrokinetic flow in a model porous material or fracture by solving numerically the Stokes--Poisson--Nernst--Planck (SPNP) equations.
In particular, we will quantify how the permeability changes as the extent of the EDL compared to channel size is varied, and we also describe how the EDL can switch the channeling of the flow in our system from regions of small aperture to regions of larger aperture.
The paper is organized as follows.
In Sec.\ \ref{sec:model}, we present the model set-up, the governing equations and their dimensionless form, in Sec.\ \ref{sec:method} we present the simulation method and our numerical scheme, and in Sec.\ \ref{sec:results} we present the results of the simulations, including validation, and effects of varying amplitude and relative Debye length.
In Sec.\ \ref{sec:discussion} we discuss technical aspects of our work and finally the conclusions and future directions follow in Sec.\ \ref{sec:conclusion}.

\section{\label{sec:model}Model}

\subsection{Flow geometry and problem set-up}
\begin{figure}[htb]
	\includegraphics[width=0.9\columnwidth]{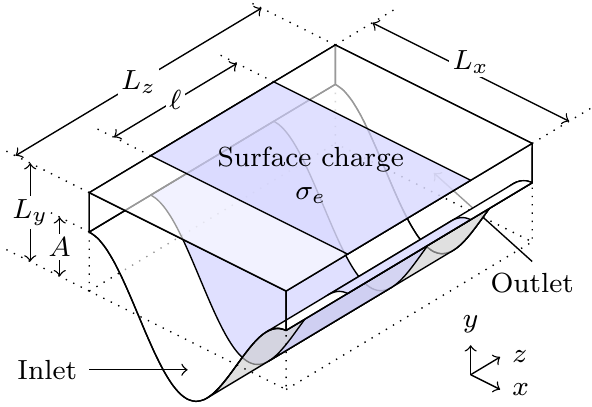}
	\caption{\label{fig:schematic}Schematic set-up of the model system. The inlet, charged-surface, and outlet areas (see text) are indicated. The $x$-direction is periodic. Note that the dimensions are not to scale.}
\end{figure}

We consider a model system consisting of an ionic solution near an undulated charged wall, as shown schematically in Fig.\ \ref{fig:schematic}.
Pressure-driven flow is imposed along the $z$-direction.
In the transverse direction, i.e.\ along the $x$-axis, the system is considered to be periodic.
In the $y$-direction the domain is bounded by two surfaces, where the bottom surface is undulated and the top surface is flat.
Along the flow direction, the domain is decomposed into three regions: an inlet region with no surface charge, a region of uniform surface charge, and an outlet region again with no surface charge.
The inlet and outlet regions must be long enough in the flow direction so as to properly account for the decay of the EDL, as discussed in more detail in Sec.\ \ref{sec:results}.

In our three-dimensional (3D) fluid slab (Fig.\ \ref{fig:schematic}), the bottom surface is described by the function $y = h(x)$ and the top surface is located at $y = L_y$.
In the plane perpendicular to $y$, our system is limited to a rectangle $(x, z) \in [0, L_x] \crossproduct [0, L_z]$.
We seek to quantify how the EDL changes our flow when the translational symmetry is broken, which we here break in the $x$-direction by a harmonic undulation,
\begin{equation}
	h(x) = A \cos \left( \frac{2 \pi x}{L_x} \right),
\end{equation}
where $A$ is the amplitude of the undulation.

\subsection{Governing equations}
The electrohydrodynamic problem is described by the SPNP equations, which couple three processes: fluid flow, electrostatics, and ion transport.
The transport of ions is described by the Nernst--Planck equation.
For ion $i$, the evolution of its number density, $n_i$, is given by:
\begin{align}
\label{eq:II:A:1}
\frac{\partial n_i}{\partial t} = \nabla \cdot \left(- n_i \mathbf{u} +D_i \nabla n_i + \frac{D_i z_i q_e }{\kB T} n_i  \nabla \varphi \right).
\end{align}
Here, $\v u$ is the fluid velocity, $D_i$ and $z_i$ are, respectively, the diffusion constant and valency for ion $i$, $q_e$ is the electron charge, $\kB$ is Boltzmann's constant, $T$ is the temperature and $\varphi$ is the electric potential. 

In the limit of negligible inertia (i.e.~Reynolds number $\mathrm{Re} \ll 1$), assuming the fluid to be incompressible, fluid flow is governed by the Stokes equations:
\begin{gather}
\label{eq:II:A:2}
\rho \frac{\partial\mathbf{u}}{\partial t} = -\nabla P + \mu \nabla^2\mathbf{u}-\rho_e \nabla \varphi, \\
\label{eq:II:A:3}
\nabla \cdot \mathbf{u} = 0.
\end{gather}
Here, $\rho$ is the density of the fluid, $P$ is the pressure, $\mu$ is the dynamic viscosity, and the charge density $\rho_e$ is given by
\begin{align}
\label{eq:II:A:4}
\rho_e = q_e \sum_{i=1}^N z_i n_i,
\end{align}
where $N$ is the number of ion species in the fluid. Finally, the electrostatic problem is given by the Poisson equation:
\begin{align}
\label{eq:II:A:5}
\nabla^2 \varphi = -\frac{\rho_e}{\epsilon_r\epsilon_0},
\end{align}
where $\epsilon_0$ is the vacuum permittivity and $\epsilon_r$ is the relative permittivity.
Together, Eqs.\ \eqref{eq:II:A:1}--\eqref{eq:II:A:5} constitute the time-dependent SPNP problem.

\subsubsection{Boundary conditions}
With regards to the velocity field, the Stokes equation is solved with a no-slip condition at the top and bottom walls of the undulated channel.
In the $z$-direction the flow is assumed to be periodic, such that the velocity field at the inlet matches that at the outlet. We drive the flow by introducing a body force along the $z$-direction, which is equivalent to having an average pressure gradient, which we denote by ${\partial P}/{\partial z}$.
In addition, we find that the resulting pressure at the inlet/outlet plane is approximately constant, and hence the solution is equivalent to having a constant-pressure boundary condition. \footnote{Our approach further prevents exceedingly long inflow and outflow transients in the coupled velocity-pressure subproblem, since the domain for the flow subproblem is closed.}

The Nernst--Planck equation is solved with a no-flux condition on the the top and bottom channel walls, and at the inlet and outlet, we prescribe the number density $n^\infty$ of the ions.
Finally, for the Poisson equation, a surface charge boundary condition is specified:
\begin{align}
\grad \varphi \cdot \hat{\vec{n}} = \frac{\sigma_e(\vec{x})}{\epsilon_r \epsilon_0},
\end{align}
where $\hat{\v n}$ is the surface normal pointing out of the domain and $\sigma_e$ is the surface charge.
The prescribed surface charge is adjusted to keep a constant surface potential, through the Grahame equation for a symmetric monovalent solution \cite{graf2006}:
\begin{align}
\label{eq:II:A:6}
\sigma_e=\sqrt{8\kB Tn^\infty\epsilon_r \epsilon_0} \sinh\left(\frac{  q_e \zeta}{2\kB T}\right).
\end{align} 
In deriving this equation, the ion-number density $n^\infty$ is considered to be set infinitely far away form the charged wall, and $\zeta$ is the surface potential when also the electric potential is set to zero at infinity.
Hence, the inlet is grounded, i.e.~$\varphi=0$, and at the outlet, a zero normal electric flux density is imposed, i.e.~$\hat{\mathbf{n}} \cdot \grad \varphi = 0$.
As indicated above, all fields are taken to be periodic along the $x$ direction.

\subsection{Dimensionless form}
For both numerical and analytical purposes, it is convenient to express the model in terms of dimensionless variables.
We further limit ourselves to a system with a symmetric monovalent ion solution, where both ions have the same diffusion constant.
\begin{table}[htb]
\caption{\label{tab:II:B:1} Physical variables, their symbols, and the normalization used in deriving dimensionless quantities, based on Ref.~\cite{nielsen2014}. Note that $n^{\infty}$ is chosen to be one of the ion number densities at the inlet.}
\begin{ruledtabular}
\begin{tabular}{l c c}
Variable & Symbol & Normalization \\
\hline
Ion number density & $n_i$ & $n^{\infty}$ \\
Electric potential & $\varphi$ & $V_T = \frac{\kB T}{z q_e}$\\ 
Length & $x$ & $R$ \\
Velocity & $\mathbf{u}$ & $U_0 = \frac{\epsilon_0 \epsilon_r V_T^2}{\mu R}$ \\
Time & $t$ & $\frac{R^2}{D}$ \\
Pressure & $P$ & $\frac{\mu U_0}{R}$ \\
\end{tabular}
\end{ruledtabular}
\end{table}
Using the scaling reported in Table \ref{tab:II:B:1} \cite{nielsen2014}, we obtain the following form of the Nernst--Planck equation \eqref{eq:II:A:1}:
\begin{align}
\label{eq:II:B:1}
\frac{\partial \tilde{n}_{\pm}}{\partial \tilde{t}} = \tilde{\nabla} \cdot \left(- \mathrm{Pe}\tilde{n}_{\pm} \tilde{\mathbf{u}} +\tilde{\nabla} \tilde{n}_{\pm}  \pm \tilde{n}_{\pm}  \tilde{\nabla} \tilde{\varphi} \right),
\end{align}
where a tilde denotes that it is a dimensionless field, and the Peclet number is defined as $\mathrm{Pe} = {R U_0}/{D}$.
Here $R$ is a typical length scale.
The Stokes equations \eqref{eq:II:A:2} and \eqref{eq:II:A:3} become:
\begin{gather}
\label{eq:II:B:2}
\frac{1}{\mathrm{Sc}} \frac{\partial\tilde{\mathbf{u}}}{\partial \tilde{t}}  = -\tilde{\nabla} \tilde{P} + \tilde{\nabla}^2\tilde{\mathbf{u}}-\frac{R^2 \kappa^2}{2}\tilde{\rho}_e\tilde{\nabla}\tilde{\varphi}, \\
\label{eq:II:B:3}
\tilde{\nabla} \cdot \tilde{\mathbf{u}} =  0
\end{gather}
where the Schmidt number is defined as $\mathrm{Sc}={\mu}/{(\rho D)}$, the Debye length is defined as 
$$\kappa^{-1} = \sqrt{\frac{\kB T \epsilon_r \epsilon_0}{2 z^2 q_e^2 n^{\infty}}},$$
and the dimensionless form for the charge density is:
\begin{align}
\label{eq:II:B:4}
\tilde{\rho}_e = \tilde{n}_+-\tilde{n}_-.
\end{align}
Finally, the Poisson equation \eqref{eq:II:A:5} becomes:
\begin{align}
\label{eq:II:B:5}
\tilde{\nabla}^2 \tilde{\varphi} = -\frac{R^2\kappa^2}{2}\tilde{\rho}_e.
\end{align}

\subsection{Time-independent form of the dimensionless equations}
In this work, we are mainly interested in the steady-state behaviour and the properties of electrohydrodynamic flow in narrow channels.
We therefore seek the time-asymptotic solutions to the coupled Eqs.\ \eqref{eq:II:B:1}, \eqref{eq:II:B:2}, \eqref{eq:II:B:3}, and \eqref{eq:II:B:5}.
The time-independent set of equations are given by:
\begin{subequations}
\label{eq:timeindependent}
\begin{equation}
\tilde{\nabla} \cdot \left(- \mathrm{Pe}\tilde{n}_{\pm} \tilde{\mathbf{u}} +\tilde{\nabla} \tilde{n}_{\pm}  \pm \tilde{n}_{\pm}  \tilde{\nabla} \tilde{\varphi} \right) = 0,
\label{eq:timeindependent_n}
\end{equation}
\begin{equation}
-\tilde{\nabla} \tilde{P} + \tilde{\nabla}^2\tilde{\mathbf{u}}-\frac{R^2 \kappa^2}{2}\tilde{\rho}_e\tilde{\nabla}\tilde{\varphi} = 0, \quad
\tilde{\nabla} \cdot \tilde{\mathbf{u}} =  0,
\label{eq:timeindependent_u}
\end{equation}
\begin{equation}
\tilde{\nabla}^2 \tilde{\varphi} = -\frac{R^2\kappa^2}{2}\tilde{\rho}_e.
\label{eq:timeindependent_phi}
\end{equation}
\end{subequations}

\section{\label{sec:method}Simulation method}

\subsection{Numerical scheme}
We solve the time-independent nonlinear equations \eqref{eq:timeindependent} equations by a splitting scheme, where the flow equations \eqref{eq:timeindependent_u} are solved in one step, while the other equations, the non-linear Poisson--Nernst--Planck (PNP) problem (Eqs.~\eqref{eq:timeindependent_n} and \eqref{eq:timeindependent_phi}), are solved in a second step using a Newton method.
The final solution is achieved by iteratively alternating between the two steps using the algorithm outlined in Ref.~\cite{mitscha-baude2017}. 
The splitting scheme results in a significant reduction in computational cost in comparison to monolithic solvers and further reduces the size of the system matrix.
Finally, the scheme permits the use of specialized solvers for the two subproblems.

\begin{algorithm}[H]
  \caption{Hybrid Solver for the SPNP system (adapted from Ref.~\cite{mitscha-baude2017}).}
  \label{alg:spnp}
   \begin{algorithmic}[1]
   \State Solve Stokes equations \eqref{eq:timeindependent_u} to obtain $(\vec{u},P)$.
   \State Solve the linearized Poisson--Boltzmann equation \eqref{eq:poissonboltzmann} to get an initial guess for $(\varphi, n_+,n_-)$.
   \State Solve one Newton step (Eq.~\eqref{eq:ApA:3}) in the PNP problem (Eqs.~\eqref{eq:timeindependent_n} and \eqref{eq:timeindependent_phi}) for $(\delta\varphi, \delta n_+,\delta n_-)$.
   \State Update $(\varphi, n_+,n_-) \leftarrow (\varphi+\delta\varphi, n_+ + \delta n_+,n_-+\delta n_-)$. 
   \State Store $(\vec{u}_{\rm old}, P_{\rm old}) \leftarrow (\vec{u}, P)$
   \State Solve Stokes equations \eqref{eq:timeindependent_u} to get ($\vec{u}$, $P$).
   \State Find $(\delta \vec{u}, \delta P) \leftarrow ( \vec{u}_{\rm old}-\vec{u}, P_{\rm old} - P)$
   \State Compute $\mathrm{Error} := \frac{1}{2}\left(\frac{ \left\Vert (\delta\varphi, \delta n_+,\delta n_-) \right\Vert}{\left\Vert(\varphi, n_+,n_-)\right\Vert}+\frac{\left\Vert (\delta \vec{u}, \delta P)\right\Vert}{\left\Vert(\vec{u},P) \right\Vert}\right)$
   \State If $\mathrm{Error} < \tau $, stop. 
   \State Else, go to Step 3 for another iteration. 
   \end{algorithmic}
\end{algorithm}

\subsection{Implementation}
Our numerical solvers are implemented in the open-source finite element framework FEniCS \cite{logg2012automated} through the Python interface to DOLFIN \cite{logg2010}.
The Stokes equation is solved using an iterative finite element solver with a pressure--convection--diffusion (PCD) preconditioner and Taylor--Hood elements, implemented in \emph{FENaPack} \cite{fenapack}.
In Appendix \ref{sec:details:num}, we derive the Newton method to solve the PNP problem.
The final method for solving the fully coupled problem is given in Algorithm \ref{alg:spnp}.
The Newton step Eq.~\eqref{eq:ApA:3} is solved using the generalized minimal residual method (GMRES) with Block Jacobi and incomplete LU preconditioning.
To achieve convergence it is essential to provide a good initial guess.
We establish an initial guess by solving the linearized Poisson--Boltzmann equation \eqref{eq:poissonboltzmann} with the same boundary conditions as the PNP problem.
Note that the preconditioning of this system is done in an \emph{ad hoc} manner and might be less robust when solving systems beyond the sizes considered here. 
More sophisticated preconditioners such as \emph{Hypre Euclid}, which was used by \cite{mitscha-baude2017} for similar purposes, were found not to be robust enough for strongly interacting EDLs.

\subsection{Mesh generation}
The mesh for the test case of a channel was generated by the built-in FEniCS function \texttt{RectangleMesh} for 2D and \texttt{BoxMesh} for 3D, that produces a structured triangular/tetrahedral mesh.
The mesh for the undulated channel was made by combining Triangle \cite{shewchuk1996}, via the Python package MeshPy \footnote{MeshPy, \url{https://documen.tician.de/meshpy/}}, to produce the surface mesh, and TetGen \cite{si2015} for the volumetric mesh. The combination of the two meshing tools allows us to produce a mesh that is periodic in both $x$- and $z$-directions.
The grid resolution  $\delta x$ was varied within the interval $\delta x \in [0.2, 0.5]$.

\section{Results}
\label{sec:results}
Using the model and methods described in the preceding sections, we performed simulations of electrohydrodynamic flow in channels in two and three dimensions, with and without undulations of the bottom surface.
The physical parameters used are given in Table \ref{tab:parameters}, although in the numerical model they enter into dimensionless quantities as given by Table \ref{tab:II:B:1} above.
The computations were performed on an in-house computing cluster using up to 28 CPU cores @ 3.0 GHz and 512 GB RAM.
\begin{table}
\caption{\label{tab:parameters} Numerical values of parameters used in the simulations, with physical units where applicable.}
\begin{ruledtabular}
\begin{tabular}{l c l l}
Quantity & Parameter  & Value & Unit \\
\hline
Ref.~concentration &$n^{\infty}$ & $[6.691-240.8]\times10^{20}$ & $ \#/\mathrm{m}^3$ \\
Zeta potential\footnote{Prescribed} &$\zeta $ & $-51.34 \times 10^{-3} $ & $\mathrm{V}$ \\ 
Channel aperture\footnote{I.e.~channel half-height.} &$a$ & $288 \times 10^{-9}$ & $\mathrm{m}$ \\
Ref.~length &$R$ & $96 \times 10^{-9} $ & $\mathrm{m}$\\
Temperature &$T$ & $298 $ & $ \mathrm{K}$\\
Diffusivity &$D$ & $1.0\times10^{-9} $ & $ \mathrm{m}^2/\mathrm{s}$\\
Boltzmann const.&$\kB $ & $1.38\times10^{-23} $ & $ \mathrm{J}/\mathrm{K}$\\
Permittivity &$\epsilon_r \epsilon_0$ & $8.854\times10^{-23} $ & $ \mathrm{C}/\mathrm{V m}$ \\
Dyn.~viscosity&$\mu$ & $1.003\times10^{-3} $ & $ \mathrm{Pa}~\mathrm{s}$\\
Electron charge&$q_e$ & $1.602\times10^{-19} $ & $ \mathrm{C}$ \\
Valency &$z$ & 1 & --\\
Pressure gradient&$\pd{P}{z}$ & $1.0\times10^{7} $ & $\mathrm{Pa}/\mathrm{m}$\\
Error tolerance &$\tau$ & $1.0\times10^{-5} $ & --
\end{tabular}
\end{ruledtabular}
\end{table}

\subsection{Electroviscous effects in a straight channel}
\label{subsec:validation}

We first validated our numerical methods against a theoretical expression for the flow in an infinitely long channel with non-interacting EDLs. 
In a straight channel (i.e.~plane Poiseuille flow between charged plates), the flow is expected to be modified from the plane Poiseuille result by an effective \emph{electric viscosity} $\mu_e$, defined through
\begin{equation}
	\langle u \rangle = \frac{a^2}{3 \mu_e}	\frac{\partial P}{\partial z}.
\end{equation}
Here, $\langle u \rangle$ is the mean velocity of the fluid, and $a$ is the channel half-height, henceforth denoted aperture.
The aperture is in physical units given by $3R$ where $R$ can be found in Table \ref{tab:parameters}.
This expression is directly related to the permeability, $K$, defined through Darcy's law by $K= \langle u \rangle \mu/ ({\partial P}/{\partial z})$.
Hence, we expect $K=\tfrac{1}{3} a^2 \mu/\mu_e$, and thus $\mu_e/\mu$ can be seen as an inverse permeability (corrected for the scaling with $a$). In our simulations, with the parameters given in Table \ref{tab:parameters}, the permeability is in the absence of electroviscous effects given by $K \simeq 28 \, \textrm{mDa}$.

In Appendix \ref{sec:details:anal}, assuming non-interacting EDLs, we derive the following theoretical estimate of the electric viscosity:
\begin{align}
\label{eq:VI:A:1}
\mu_e = \mu\left[ 1 - \frac{6\beta}{\kappa^2a^2\Fcc}f\left(\kappa a,\beta\right) \left(1 - \frac{1 }{\kappa a } \tanh(\kappa a) \right)\right]^{-1},
\end{align}
where $\beta = {\epsilon_r\epsilon_0\zeta^2}/({\mu D})$, and $\zeta$ is the surface potential, and we have used the expressions
\begin{gather}
\label{eq:VI:A:2}
\Fcc = \int_{0}^{1}2\cosh\left(\frac{q_e \zeta}{\kB T }\frac{\cosh(\kappa a X)}{\cosh(\kappa a)}\right)\mathrm{d}X,\\
\label{eq:VI:A:3}
f\left(\kappa a,\beta\right)=\frac{1 -  \frac{1}{\kappa a}\tanh(\kappa a) }{1+\frac{\beta}{\Fcc}\left(\frac{1}{\kappa a}\tanh(\kappa a)-\mathrm{sech}^2(\kappa a) \right)}.
\end{gather}
The integral in the expression for $\Fcc$ is computed numerically. 

The expression for the streaming potential $\Vstr$ is given by 
\begin{align}
\label{eq:VI:A:4}
\Vstr = \frac{2 \zeta}{\mu D \kappa^2 \Fcc}f ( \kappa a,\beta ) \Delta P.
\end{align}

We compare our simulations to the analytical prediction of $\mu_e$ by integrating the total fluid flux 
$Q_{z} (\kappa a)$
through the channel for a range of values of the ratio of aperture to Debye length, $\kappa a$. 
Note that we vary $\kappa a$ indirectly, by varying $n^\infty$.
Then, we use the following relation:
\begin{align}
\label{eq:VI:A:5}
\frac{\mu_{e,h}}{\mu} = \frac{Q_{z}(0)}{Q_{z}(\kappa a)}.
\end{align}
As only half of the length of the channel in our numerical simulations is charged, we denote the resulting electro-viscosity by $\mu_{e,h}$. 
In order to obtain a value for the electric viscosity that should correspond to the theoretical one, we scale it in the following way:
\begin{align}
\frac{\mu_e}{\mu} =  \frac{\frac{\mu_{e,h}}{\mu}-1}{\frac{l}{L_z}} +1,
\label{eq:mu_e_trans}
\end{align}
The value in the denominator is the ratio of the length of the charged part of the channel, $l$, to total length, $L_z$, such that in our simulations we have that $l/L_z=0.5$.
As increased dissipation is expected mainly to take place in the charged part of the channel, we have here ignored inlet and outlet effects, and the accuracy of these expressions would therefore improve for longer domains.
The streaming potential is measured by: 
\begin{align}
\frac{\Vstr}{l} = \frac{\int_{\Gamma_{\rm outlet}} \varphi \mathrm{d} \Gamma }{l\int_{\Gamma_{\rm outlet}}  \mathrm{d} \Gamma }
\end{align}
where the integral is taken over the outlet boundary of the domain, $\Gamma_{\rm outlet}$.

\begin{figure}[htb]
	\includegraphics[width=0.95\columnwidth]{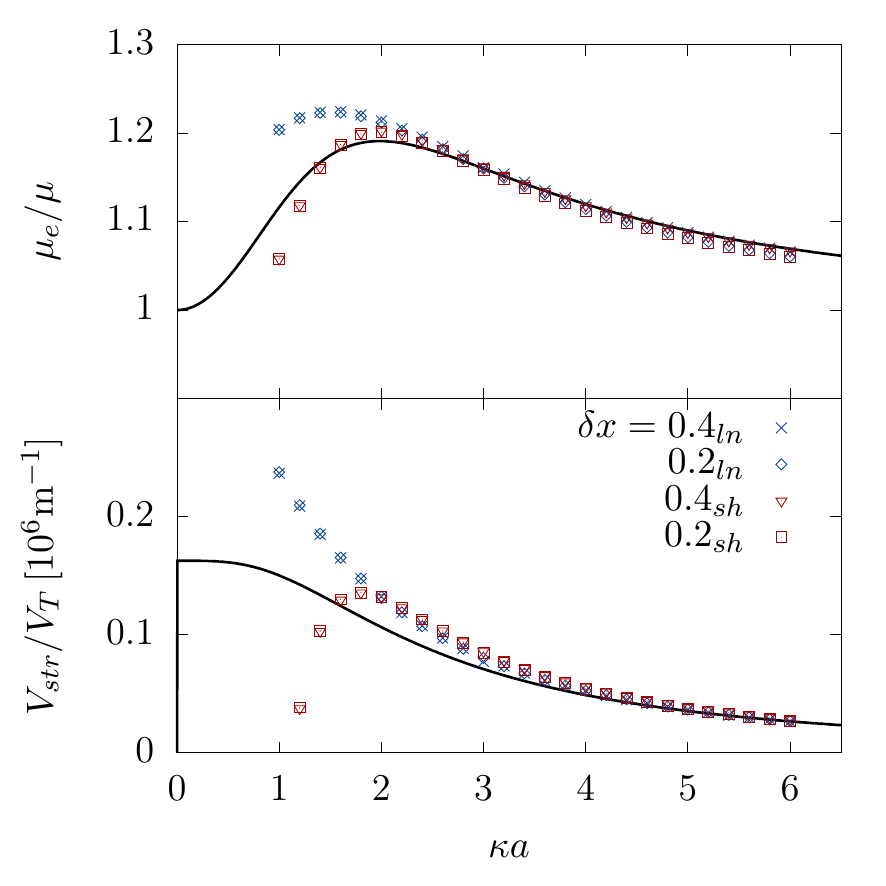}
	\caption{\label{fig:2dconvegens}
	Comparison of 2D simulations of two channel lengths to theoretical predictions.
    The blue points, where the resolution $\delta x$ has a subscript $ln$ correspond to simulations with channel length $160 R$, and the red points, with a subscript $sh$, to simulations with channel length $40 R$.
    Both channels have a width of $6 R$.
    The solid lines denote the analytical results from Eqs.~\ref{eq:VI:A:1} (top) and \ref{eq:VI:A:4} (bottom).
    In both cases the center half of the channel has a surface charge.
    Top: Plot of the electric viscosity as a function of $\kappa a$.
    Bottom: The streaming potential in units of the thermal voltage.}
\end{figure}

We tested our numerical simulations against the analytic results using both 2D and 3D versions of our code. In addition, we tested the influence of the numerical resolution on the results. In Fig.~\ref{fig:2dconvegens}, we present plots of the measured electric viscosity (top panel) and the streaming potential per length (bottom panel) for 2D simulations, compared to the theoretical predictions of, respectively, Eqs.~\eqref{eq:VI:A:1} and \eqref{eq:VI:A:4}. For the theoretical curves, the $\zeta$ potential in Eq.~\eqref{eq:II:A:6} is not used directly. Instead we use an empirical value computed from our simulations, which here in physical units has the value $\zeta = -45.2 \mathrm{mV}$. The quantities are plotted as function of $\kappa a$, i.e.~the ratio of the channel aperture to the Debye length. We also investigate the effect of the domain length using two lengths, $L_x = 40 R$ and $L_x=160 R$. With regards to the electric viscosity, shown in the top panel of \ref{fig:2dconvegens}, it is clear that the value $\mu_e$ approaches the theoretical one for large values of $\kappa a$ but departs for small values of $\kappa a$.
This departure has different reasons for the two channel lengths.
For the long channel, the departure arises because the linear Poisson--Boltzmann theory breaks down when we have strongly interacting EDLs, and for the short channel the departure is caused by surface charge which cannot be screened within the domain.
The effect of strongly overlapping EDLs could be incorporated into the theoretical estimate by solving the non-linear Poisson--Boltzmann equation numerically, or using the implicit solution found by \citet[pp.~67]{verwey1948} and extending the procedure in Appendix \ref{sec:details:anal}.

In the bottom panel of Fig.~\ref{fig:2dconvegens}, we observe that the streaming potential $\Vstr$ is in good agreement with the theory in the limit of large values of $\kappa a$. The departure from the theoretical prediction for small values of $\kappa a$ appears for the same reasons as for the electric viscosity.


We further validated that our 2D steady-state solver gives the \emph{physically} correct solution by comparing with the asymptotic solution to the full time-dependent equation system. For that purpose, we applied the independently developed time-dependent solver \cite{linga2018a}, implemented in the \emph{Bernaise} framework \cite{linga2018b}, for flow through a circular packing with similar boundary conditions as considered in this paper.
It was confirmed that the time-dependent solver approached the steady-state solution in the large-time limit; in particular, the difference in streaming potential was less than $1\%$ after a simulation time $T \simeq 5 \tau_D$, where the Debye length based diffusive time scale is $\tau_D = \kappa^{-2} / D = 1.5^2 \cdot 2.189 \simeq 5$ (see \cite{linga2018a} or Supplementary Material).

\begin{figure}[htb]
	\includegraphics[width=0.95\columnwidth]{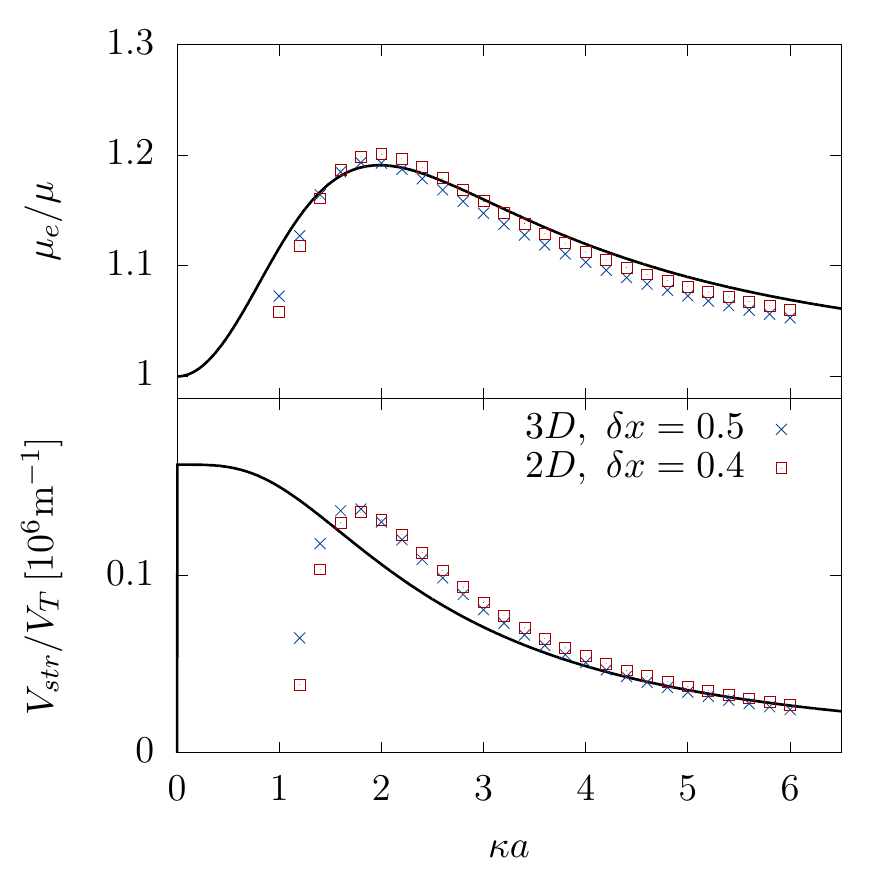}
	\caption{\label{fig:2dvs3d} Comparison of 3D versus 2D channel flow simulations.
    The 2D channel is the short channel described in Fig.~\ref{fig:2dconvegens}, and the 3D channel has the same size in the streamwise and vertical dimensions, while the additional horizontal dimension is periodic with length $R$.
    The analytical predictions shown as solid lines are the same as in Fig.~\ref{fig:2dconvegens}.
    Top: Electric viscosity as a function of $\kappa a$.
    Bottom: The streaming potential in units of thermal voltage as a function of $\kappa a$.}
\end{figure}

Fig.~\ref{fig:2dvs3d} shows a comparison of our 3D simulations with 2D simulations in equivalent geometries, i.e.~geometries translationally invariant in the transverse directon.
In the top panel, we see that the curves for the electric viscosity coincide, meaning that the 3D simulations give comparable results to the 2D case.
Likewise, we see in the bottom panel that the streaming potentials of 2D and 3D compare well to each other.
This gives a strong indication that the full 3D simulation constitutes a reliable approach.

An apparent discrepancy between the analytical and simulated results occurs when the Debye length becomes larger than the channel height.
There are two reasons for this; (i) the overlapping double layers from top and bottom, and (ii) leakage of ions into to the inlet and outlet zones, which leads to unphysical boundary conditions and a spurious negative streaming potential.
The latter effect can be compensated by extending the inlet and outlet zones to be sufficiently long, such that to a good approximation, \emph{both} $\tilde{n}_\pm = 1$ and $\hat{\mathbf{n}} \cdot \grad \tilde{V} = 0$ at both inlet and outlet \footnote{Note that only the former is enforced at the inlet boundary, and both are enforced at the outlet boundary.}.

\subsection{Macroscopic effects due to an undulated surface}
\begin{figure}[htb]
	\includegraphics[width=0.95\columnwidth]{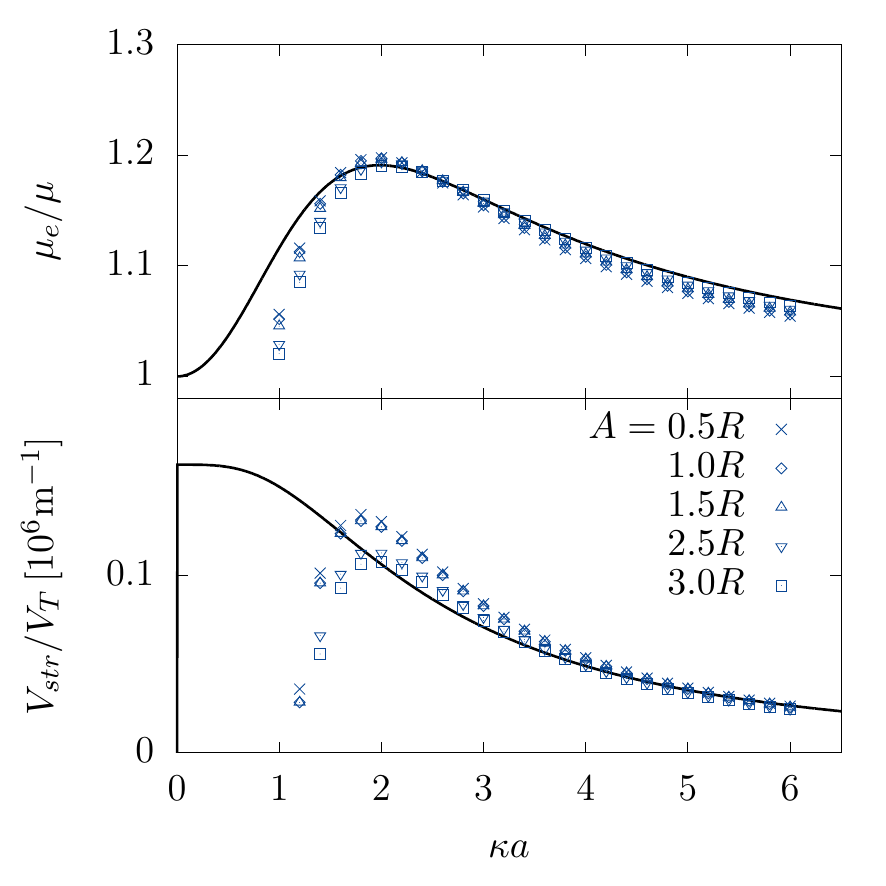}
	\caption{\label{fig:mu_eandV_strwave}
    	Comparison of electric viscosity and streaming potential in 3D channels with varying undulation amplitude.
        The channels, shown schematically Fig.~\ref{fig:schematic}, have dimensions given in the text, and $A$ is given in the legend.
        The solid lines are theoretical predictions and the same as in Figs.~\ref{fig:2dconvegens} and \ref{fig:2dvs3d}.
        Top: The electric viscosity plotted as a function of $\kappa a$.
        Bottom: The streaming potential in units thermal voltage plotted as a function of $\kappa a$.}
\end{figure}

In order to quantify how the flow is affected by electro-viscous effects in uneven channels, simulations were run in the geometry shown in Fig.~\ref{fig:schematic} with an amplitude $A$  varying from $0.5R$ to $3R$, and the other dimensions fixed to $L_x =12 R, L_y = 6 R, L_z = 40 R$, and $l=20R$.
From these simulations, we calculated $\mu_e$ and $\Vstr$ as described in section \ref{subsec:validation}, and the results are shown in Fig.~\ref{fig:mu_eandV_strwave}.
As shown in the top panel, the electric viscosity does not seem to be strongly affected, but it is worth noting that for increasing amplitude, a slight decrease is observed for small values of $\kappa a$.
The streaming potential, shown in the bottom panel, seems to be more affected by the change of amplitude.
This could be due to the overlap of double layers in the narrow regions, leading to a stronger non-linear effect, but also more leaking to the boundary (i.e.~a finite-size effect).
However, these plots yield limited insight into the effect of any asymmetry induced by the undulation, as these quantities are averaged over the whole domain.

\subsection{Quantification of flow channeling}
In order to quantify the asymmetry induced by the electro-viscous effect in the charged part of the channel, we define the following subdomains of $\Omega$:
\begin{align}
\label{eq:VI:B:1}
\Omega_{t} &= [0,12R] \crossproduct [-3R,6R] \crossproduct [15R,25R],\\
\label{eq:VI:B:2}
\Omega_{y} &= [0,12R] \crossproduct [3R,6R] \crossproduct [15R,25R],\\
\label{eq:VI:B:3}
\Omega_{x} &= [-3R,3R] \crossproduct [-3R,6R] \crossproduct [15R,25R].
\end{align}
Note that the domain of $\Omega_{y}$ has half the volume within the computational domain compared to that of $\Omega_{t}$ for the undulated channel, as long as the amplitude is smaller or equal to $3$R.
We then integrate the longitudinal component of the velocity field, $u_z$, in the subdomains and divide by the length in order to find the average flux through each subdomain:
\begin{align}
Q_{z,i} (A,\kappa a) = \frac{1}{10R}\int_{\Omega_i} u_z \, \mathrm{d}v, \quad i\in \{t,x,y\}.
\label{eq:def_QzA}
\end{align}
Now, we define the absolute asymmetries $\Theta_{x}$ and $\Theta_{y}$ by
\begin{align}
\Theta_{i} (A,\kappa a) = \frac{Q_{z,i} (A,\kappa a)}{Q_{z,t} (A,\kappa a)}, \quad i\in \{x,y\},
\label{eq:def_ThetaA}
\end{align}
and finally the relative asymmetries $\theta_x$ and $\theta_y$ by
\begin{align}
\theta_{i} (A,\kappa a) = \frac{\Theta_{i}(A,\kappa a)}{\Theta_{i}(A,0)}, \quad i\in \{x,y\}.
\end{align}
This quantity gives a measure of how the flow is re-distributed between regions of small and large aperture ($\theta_x$) and between top and bottom ($\theta_y$) due to a surface undulation, with amplitude $A$, and the effect of EDL, through $\kappa a$.

\begin{figure}[htb]
	\includegraphics[width=0.8\columnwidth]{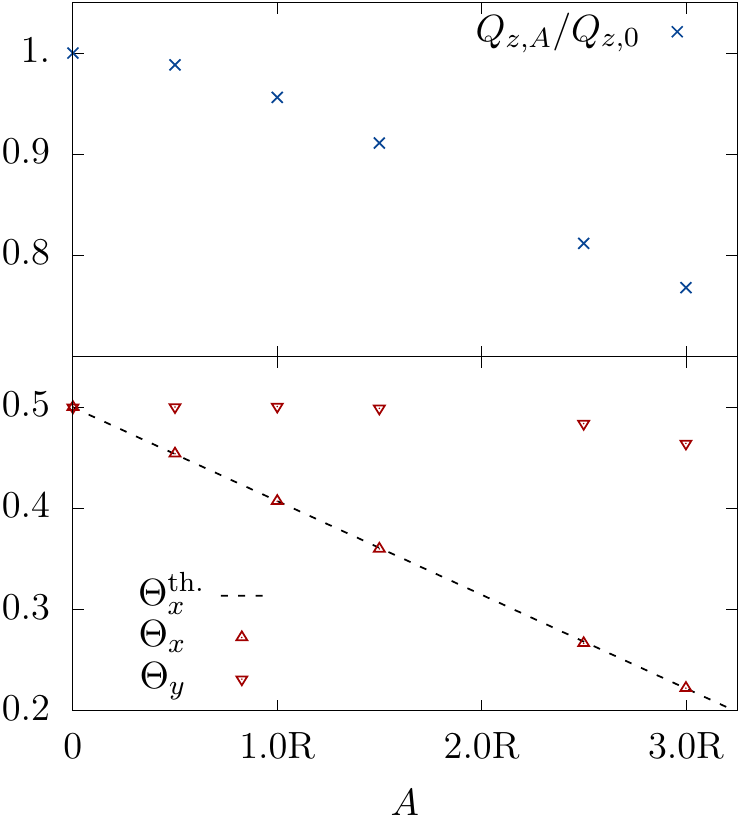}
    \caption{\label{fig:flowrat}
    Simulations in an undulated 3D channel without any electric effects included, i.e.~$\kappa a =0$.
    Top: Flow rate in undulated channels as a function of undulation amplitude $A$, relative to the flat channel $A=0$.
    Bottom: The absolute asymmetry of the flow in the channel as a function of amplitude $A$.
    The linear dependence of $\Theta_x$ on $A$ is in good agreement with the theoretical prediction $\Theta_x^{th.}$ derived in Appendix \ref{sec:smallamp}.}
\end{figure}

It is interesting to first consider the isolated effect of an undulated geometry, i.e.~flow without any electric effects, but with a variable amplitude on one of the sides.
This is achieved by setting $\kappa  a = 0$ in our simulations.
In principle, this limiting case results in translational symmetry along the streamwise direction and thus reduces to a 2D Poisson problem (see Appendix \ref{sec:smallamp}), but here we show results from full 3D simulations.
In Fig.~\ref{fig:flowrat}, both the relative flow rate, $Q(A, 0)/Q(0, 0)$, and the absolute asymmetries, $\Theta_i(A, 0)$, are plotted as a function of amplitude $A$.
In the top panel, we see that the total flow rate is significantly reduced.
In the bottom panel, the absolute asymmetry along the vertical direction displays a rather weak dependence on the amplitude (it becomes pronounced only at $A=2.5R$), while the absolute asymmetry along the $x$ direction seems to depend linearly on the amplitude.
This is also in agreement with the theoretical prediction based on a first-order expansion in the undulation amplitude $A$ obtained in Appendix \ref{sec:smallamp}.

\begin{figure}[htb]
	\includegraphics[width=0.95\columnwidth]{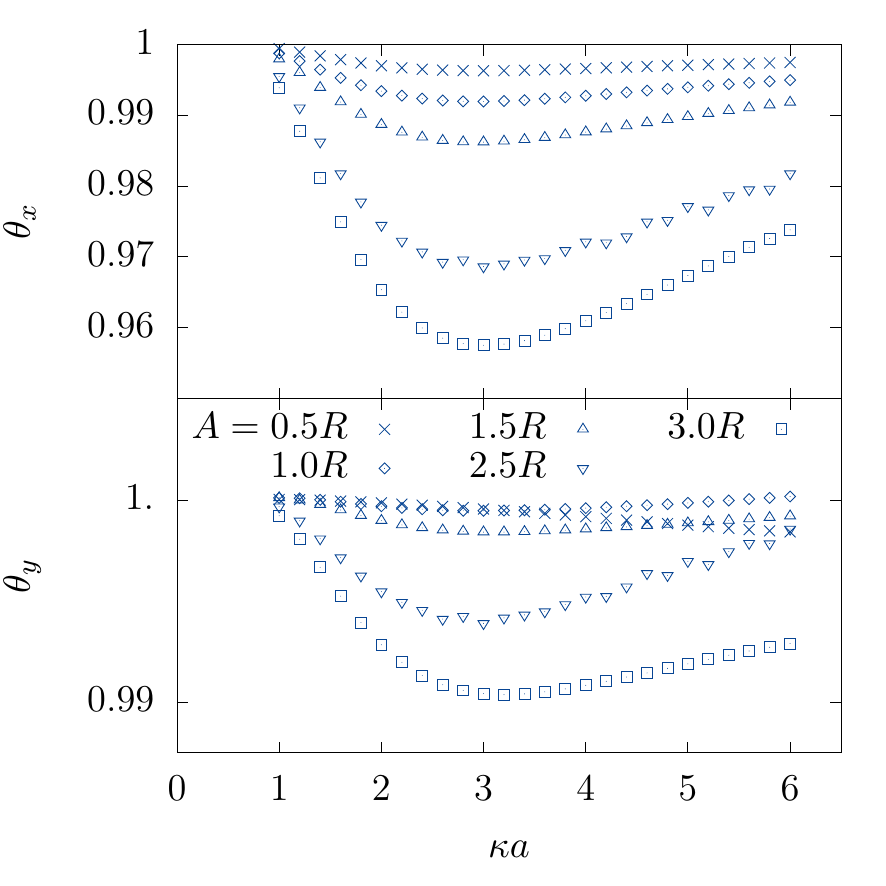}
	\caption{\label{fig:asymmetry}
    Relative asymmetries are plotted as a function of $\kappa a$.
    Top: The relative asymmetry in the plane of flow, $\theta_x$.
    Bottom: The relative asymmetry normal to the plane of flow, $\theta_y$.
    }
\end{figure}

The plots presented in the bottom panel of Fig.~\ref{fig:flowrat}, without electric effects, serve as a reference for the simulations with electrohydrodynamic effects, i.e.~the relative asymmetries $\theta_i$  for $\kappa a > 0$.
Plots of the relative asymmetries $\theta_{x} (A,\kappa a)$ and $\theta_{y} (A,\kappa a)$ are shown in Fig.~\ref{fig:asymmetry}.
Inspecting $\theta_x$ in the top panel of Fig.~\ref{fig:asymmetry}, we see that there is an increased damping of the flow in the narrow part of the channel, which means that the electric effects amplifies the channeling beyond what is caused by the amplitude alone (shown in Fig.~\ref{fig:flowrat}, bottom panel).
The effect in the vertical direction is weaker, as shown in the bottom panel of Fig.~\ref{fig:asymmetry}, and only becomes visible when the amplitude is large enough to form a narrow region in the bottom of the channel.
Even then, the effect is less than $1 \%$.
It should be noted that in our simulations only half of the channel is charged, and if a larger fraction of the wall was charged, the effect would presumably be stronger.
An open question is whether this effect is linear in the length of the charged domain, as is the case with the electric viscosity, see Eq.\ \eqref{eq:mu_e_trans}.

\subsection{Local effects}
In order to get a detailed understanding of the increased asymmetry and channeling of the flow in the undulated channel, we visualize the local ratio between the flow field with and without electric effects. To this end, we measured $u_z$ in 40 cross sections evenly spaced in the interval $z\in[12.5R, 27.5R]$, which is inside the charged region of the channel. This was done for both the uncharged realization and the one 
corresponding to $\kappa a = 3.0$. The 40 cross sections were averaged in order to cancel out noise, and we denote the resulting $z$-averaged fields by $\langle u_z \rangle_z (\kappa a)$.
However, near the walls there are still some artifacts present (see Fig.~\ref{fig:flowfield} below) due to a structured surface mesh and an amplification of errors as the reference solution was near 0 here---a consequence of the no-slip condition. In Fig.~\ref{fig:flowfield}, we plot the ratio between the charged and non-charged flow fields, $\langle u_z \rangle_z (\kappa a)/\langle u_z \rangle_z (0)$. The panels in the figure show increasing amplitudes $A$ for a fixed $\kappa a = 3$. In panels a) and b), we see that the main difference is in the boundary layer near the walls and there is only a minor increase in channeling.
For sufficiently high amplitudes, shown in panels c) and d), it is clear that the flow is channeled to the region where the amplitude is largest. 
The local change in the flow rate is of the order of $10$--$15 \%$ in the narrow regions, particularly visible near the walls.
\begin{figure}[htb]
	\includegraphics[width=0.9\columnwidth]{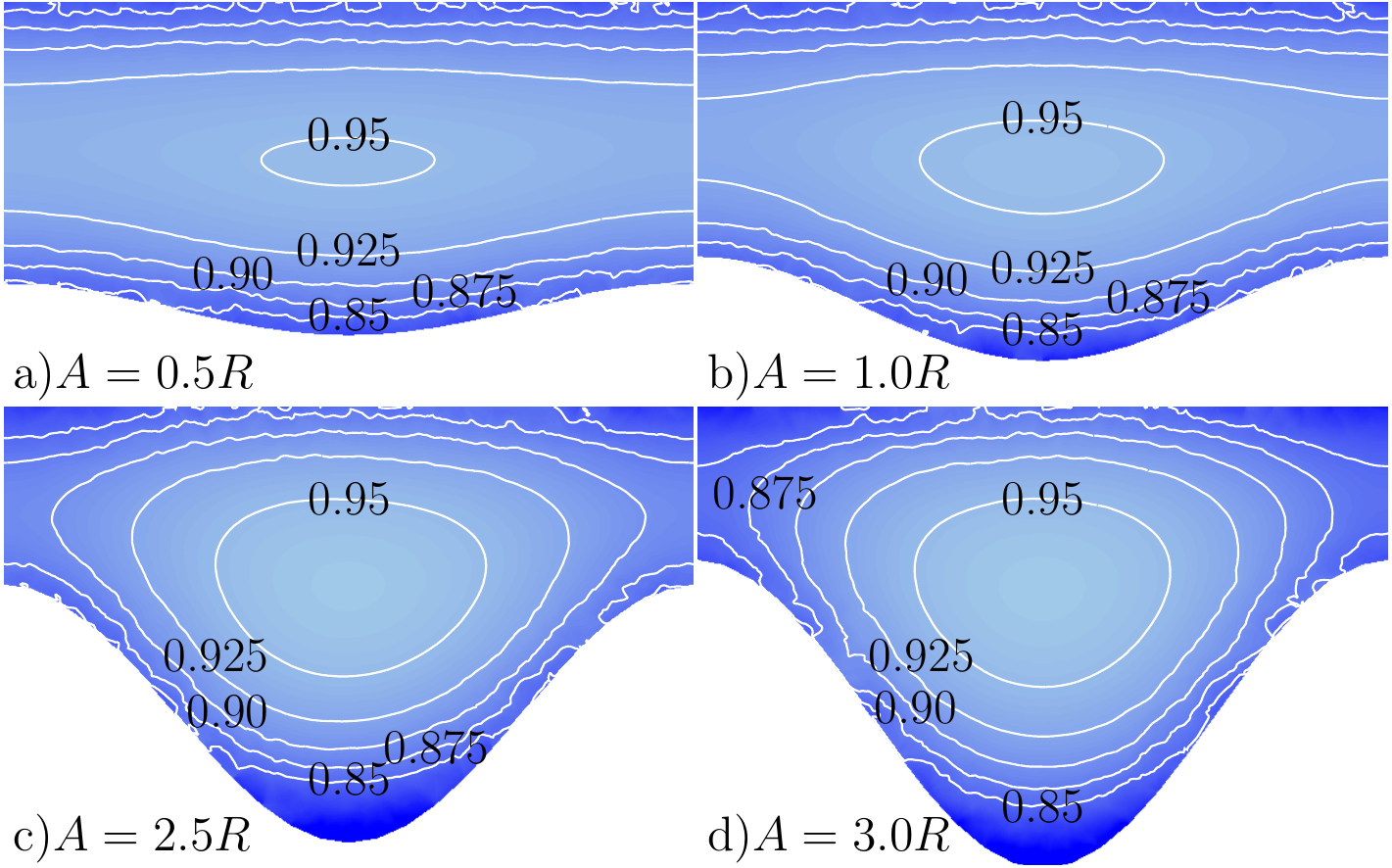}
	\caption{\label{fig:flowfield}
    The flow field at  $\kappa a = 3$ divided by the flow field for the flow without any electric effects, probed as described in the text. Figures a)--b) show increasing amplitude.}
\end{figure}

\section{Technical discussion}
\label{sec:discussion}

In Figs.~\ref{fig:2dconvegens}--\ref{fig:mu_eandV_strwave} and \ref{fig:asymmetry}, we have plotted the physical quantities as a function of the ratio,  $\kappa a$, of the channel aperture and the Debye length. We note that there is a subtelety when varying $\kappa a$, either through the Debye length $\kappa^{-1}$ or by tuning the channel aperture $a$. This is due to the quadratic dependence on $\kappa$ in Eq.~\eqref{eq:VI:A:4}.
In this work, we have held $a$ fixed while varying $\kappa$, indirectly by setting the reference concentration $n^\infty$.
There is also another effect at play when approaching low concentration in finite channels, namely that the equilibrium approach of the Poisson--Boltzmann equation is less accurate as the advection term becomes more dominant in the Nernst--Planck equation. 
This effect could in part be responsible for the increase in the streaming potential, which was observed from Fig.~\ref{fig:2dconvegens} for the long tube.
\citet{mansouri2005} avoided such complications by instead varying $a$ when addressing the dependence of the streaming potential on $\kappa a$ for axially symmetric capillaries.

Here, we have only simulated dilute solutions. When the ion concentrations approach that of e.g.~sea water, we would have to include the effect of dispersion forces near the charged walls, make high density corrections to the Nernst--Planck chemical potential as well as take into account other strongly coupled phenomena \cite{bostrom2001}.  
 
As pointed out in Sec.~\ref{sec:results}, there is a quite pronounced effect of having a short inlet and outlet. This was however necessary in order to run full simulations in 3D since the PNP problem becomes increasingly hard to solve numerically when the system size increases -- a hallmark of ill-preconditioned matrices for Krylov-subspace solvers.
Therefore, in order to handle larger systems, we would have to either rely on a direct solver implying a massive increase in the need of computational resources, or to find a better preconditioner.
This would allow a deeper investigation of the regime where the EDLs overlap and the linear Poisson--Boltzmann theory breaks down.

\section{\label{sec:conclusion}Conclusion}
Flow in highly irregular geometries with charged surfaces is commonplace in many geological and industrial settings. In some situations, even a moderate change of the local flow distribution can have an impact on the precipitation and chemical reactions \cite{steefel2005}. We have in this paper considered the electrohydrodynamic effects on flow by numerically solving the Stokes--Poisson--Nernst--Planck equation in narrow undulated channels.
The undulated channel geometry serve as a simplified model of micro-scale fractures, which often mediate the large-scale transport e.g.\ in porous rock.
By varying the amplitude of the channel undulation and the Debye length, we have analyzed the macroscopic flow changes in terms of the streaming potential and electric viscosity. Further, we have observed an enhanced channeling of the flow. In particular, we observe for the larger undulation amplitudes up to  $5 \%$ flux reductions, relative to a system without surface charge. The local flow may vary as much as $10 \%$.
In comparison to pure hydrodynamic channeling, our results indicate that ridges may be even more prone to precipitation than valleys, leading to a positive feedback with enhanced channeling effects.

Our results offer insight into electrodydrodynamic flow in realistic pore and fracture geometries. 
Further studies would be of interest, primarily in larger and more complex samples,
to get an even deeper understanding of electrohydrodynamic effects in geological settings. Further, it would be interesting to study the precipitation and/or dissolution dynamics in the presence of surface charge.
Finally, electrohydrodynamics might be important in two-phase flow, where the local forces could alter the wetting properties and hence control the macroscopic fluid flow.

\begin{acknowledgments}
This project has received funding from the Villum Foundation through the grant ``Earth Patterns'', and from the European Union's Horizon 2020 research and innovation program through Marie Curie initial training networks under grant agreement 642976 (NanoHeal).
The authors are thankful to Henrik Bruus and Fran\c{c}ois Renard for stimulating discussions.
\end{acknowledgments}

\appendix

\section{Newton method for the Nernst--Planck--Poisson problem}
\label{sec:details:num}
The Nernst--Planck--Poisson problem takes the following non-linear weak form: 
\begin{align}
\label{eq:ApA:1}
0 = \int_{\Omega}^{}  &\psi \laplacian \varphi + \psi \frac{R^2\kappa^2}{2}\left(n_+-n_-\right)\\\nonumber 
&-c_+{\mathrm{Pe}}\div\left(\mathbf{u}n_+\right)+ c_+\laplacian n_+ +c_+\div\left(n_+\grad\varphi\right)\\\nonumber
&-c_-{\mathrm{Pe}}\div\left(\mathbf{u}n_-\right)+ c_-\laplacian n_- -c_-\div \left(n_-\grad\varphi\right)\mathrm{d}v,
\end{align}
where $\psi$ is the test function for the electric potential and $c_+, c_-$ are the test functions for the cation and anion number densities respectively.
We can develop a Newton method for solving the equation by viewing the weak form in Eq.~\eqref{eq:ApA:1} as a functional called  $F(\mathbf{U})$, where $\mathbf{U}=(\varphi,n_+,n_-)$, and then expanding around some $\mathbf{U}^0$.
This gives:
\begin{align}
\label{eq:ApA:2}
0=F(\mathbf{U}^0) + \left.\int_{\Omega}^{}\frac{\delta F(\mathbf{U})}{\delta \mathbf{U}}\delta\mathbf{U}\mathrm{d}v\right\rvert_{\mathbf{U}=\mathbf{U}^0} + \mathcal{O}(\delta^2),
\end{align}
where $\delta \mathbf{U}$ is a variation away from $\mathbf{U}^0$. 
Now, performing this for Eq.~\ref{eq:ApA:1} and applying the appropriate boundary conditions gives the following linearized weak form:
\begin{multline}
\label{eq:ApA:3}
0 = \int_{\Omega}^{} \Big[ -\grad\psi \cdot \grad\varphi^0 
+ \psi \frac{R^2\kappa^2}{2}\left(n_+^0-n_-^0\right) \\
+\frac{1}{\mathrm{Pe}}\grad c_+\cdot\left(\mathbf{u}n_+^0\right)
-\grad c_+\cdot\grad n_+^0 
- \grad c_+\cdot\left(n_+^0\grad\varphi^0\right) \\
+\frac{1}{\mathrm{Pe}}\grad c_-\cdot\left(\mathbf{u}n_-^0\right)
- \grad c_-\cdot\grad n_-^0 
+\grad c_-\cdot\left(n_-^0\grad\varphi^0 \right) \Big] \mathrm{d}v \\
+ \int_{\Omega}^{} \Big[ 
-\grad\psi \grad \delta\varphi 
+ \psi \frac{R^2\kappa^2}{2}\left(\delta n_+-\delta n_-\right) \\ 
+\frac{1}{\mathrm{Pe}}\grad\cdot c_+\left(\mathbf{u} \delta n_+\right)
- \grad c_+\cdot\grad \delta n_+ \\
-\grad\cdot c_+\left( \delta n_+\grad\varphi^0\right)
-\grad\cdot c_+\left(  n_+^0\grad\delta\varphi\right) \\
+\frac{1}{\mathrm{Pe}}\grad c_-\cdot\left(\mathbf{u}\delta n_-\right)
- \nabla c_-\cdot\grad\delta n_- \\ 
+\grad c_-\cdot\left( \delta n_-\grad\varphi^0\right) 
+\grad c_-\cdot\left( n_-^0\grad\delta\varphi\right) \Big] \mathrm{d}v \\
+ \int_{\Gamma}^{}  \psi\frac{R^2\kappa^2}{2}\sigma_e \mathrm{d}s.
\end{multline}
This weak form can then been discretized and solved using the finite element method.

\section{Analytical expressions for the electroviscous effect}
\label{sec:details:anal}
Here we derive analytical expressions for flow in a channel, used as comparison to our numerical simulations.
The derivation follows closely the one found in \cite{rice1965,mansouri2005}, but are considered with channel flow instead of flow in a tube.

Consider the steady state Nernst--Planck equation with a zero velocity field: 
\begin{align}
\label{eq:ApB:1}
\div(D_in_i \grad g_i) = 0,
\end{align}
where $g_i$ is the chemical potential defined as:
\begin{align}
\label{eq:ApB:2}
g_i = \ln(n_i) +\frac{q_e z_i}{k_b T}\varphi 
\end{align}
now if Eqs.~\ref{eq:ApB:1} has to be satisfied, $n_i$ must be given by: 
\begin{align}
\label{eq:ApB:3}
n_i = n^\infty\exp\left(-\frac{q_e z_i}{k_b T}\varphi\right),
\end{align}
where $n^\infty$ is the mean/inlet number density of the ions. 
Plugging Eqs.~\ref{eq:ApB:3} for a symmetric mono-valent solution into Eqs.~\ref{eq:II:A:5} yields the Poisson--Boltzmann equation:
\begin{align}
\label{eq:ApB:4}
\laplacian \varphi = \frac{2 q_e n^\infty}{\epsilon_r \epsilon_0} \mathrm{sinh} \left(\frac{q_e z_i}{k_bT}\varphi\right).
\end{align}
Now expanding in $\varphi$ around zero linearizes the Poisson--Boltzmann equation:
\begin{align}
\label{eq:ApB:5}
\laplacian \varphi = \kappa^2 \varphi.
\end{align}
This can now be solved in a channel with walls at $x=\pm a$, under the following boundary conditions $\left.\varphi\right\rvert_{x=\pm a} = \zeta$, and that its transverse derivative is zero in the center of the channel, $\left.\pd{\varphi}{x} \right\rvert_{x=0} = 0 $.
This gives: 
\begin{align}
\label{eq:ApB:6}
\varphi(x)=\zeta\frac{\cosh(\kappa x)}{\cosh(\kappa a)}, 
\end{align}
and the charge density with in the linear approximation becomes: 
\begin{align}
\label{eq:ApB:7}
\rho_e(x)=-\zeta \epsilon_r \epsilon_0 \kappa^2 \frac{\cosh(\kappa x)}{\cosh(\kappa a)}. 
\end{align}
Now consider the Stokes equation in the same infinitely long channel with a pressure gradient and electric field along the $z$-direction: 
\begin{align}
\label{eq:ApB:8}
\mu\ \pdd{u_z}{x} = - \pd{ P}{z} - \rho_e E_z, 
\end{align}
Solving this with a no-slip condition at the walls and the charge density for Eqs. \ref{eq:ApB:7} yields 
\begin{align}
\label{eq:ApB:9}
u_z(x) = \frac{\pd{ P}{z}}{2 \mu}\left(a^2-x^2\right) - \frac{\epsilon_r \epsilon_0 \zeta E_z}{\mu} \left(1 - \frac{\cosh(\kappa x)}{\cosh(\kappa a)}\right).
\end{align}
Now, to close the system, we assume that the charge-current-flux along the $z$ direction in the channel vanishes at steady state.
The charge-current-density is given as:
\begin{align}
\label{eq:ApB:10}
J_{e(z)} = \rho_e(x) u_z(x) +\frac{2 D q_e^2}{\kB T }E_z n^\infty \cosh\left(\frac{q_e \varphi(x)}{\kB T }\right),
\end{align}
and integrating it over the channel cross section gives the flux:
\begin{align}
\label{eq:ApB:11}
\mathcal{L}\int^{a}_{-a} J_{e(z)} \mathrm{d}x = &- 2\mathcal{L} a \frac{\partial P}{\partial z} \Omega\left[1 -  \frac{1}{\kappa a }\tanh(\kappa a) \right] \\\nonumber 
&+ {\mathcal{L}a \Omega^2 E_z \mu \kappa^2}\left[\frac{1}{\kappa  a}\tanh(\kappa a)-\mathrm{sech}^2(\kappa a) \right] \\\nonumber
&-\frac{2 \mathcal{L} a D q_e^2n^\infty}{\kB T }E_z \Fcc,
\end{align}
where $\Omega=\frac{\epsilon_r \epsilon_0 \zeta}{\mu}$ and 
\begin{align}
\label{eq:ApB:12}
\Fcc = \int_{0}^{1}2\cosh\left(\frac{q_e \zeta}{\kB T }\frac{\cosh(\kappa a X)}{\cosh(\kappa a)}\right)\mathrm{d}X.
\end{align}
Now using the the no-flux condition to get an expression of the ration of $E_z$ and $\pd{P}{z}$
gives:
\begin{align}
\label{eq:ApB:13}
\left.\frac{E_z}{\pd{P}{z}}\right\rvert_{J_{e(z)}=0}= \frac{2 \zeta }{\mu D \kappa^2 \Fcc}f\left(\kappa a,\beta\right),
\end{align}
where 
\begin{gather}
\label{eq:ApB:14}
f\left(\kappa a,\beta\right)=\frac{1 -  \frac{1}{ a\kappa }\tanh(\kappa a) }{1+\frac{\beta}{\Fcc}\left(\frac{1}{a\kappa}\tanh(\kappa a)-\mathrm{sech}^2(\kappa a) \right)},\\
\label{eq:ApB:15}
\beta = \frac{\epsilon_r\epsilon_0 \zeta^2}{\mu D}.
\end{gather}
Integrating Eqs.~\ref{eq:ApB:13} from one end of the channel to the other gives the Helmholtz--Smoluchowski equation:
\begin{align}
\label{eq:ApB:16}
\Vstr = \frac{2 \zeta}{\mu D \kappa^2 \Fcc}f ( \kappa a,\beta ) \Delta P.
\end{align}
To find the electroviscous effect, we use Eqs.~\ref{eq:ApB:13} to eliminate $E_z$ in Eqs.~\ref{eq:ApB:9} and integrate to get the velocity flux $Q_z$ along the $z$-direction: 
\begin{align}
\label{eq:ApB:17}
	Q_z = \frac{2 \mathcal{L} a\frac{\partial P}{\partial z}}{3 \mu}\left[ a^2 - \frac{6\beta }{\kappa^2 \Fcc}f\left(\kappa a,\beta\right) \left(1 - \frac{1}{\kappa a } \tanh(\kappa a) \right)\right].
\end{align}
Now, the electric viscosity $\mu_e$ must be defined implicitly by: 
\begin{align}
\label{eq:ApB:18}
Q_z=\frac{2 \mathcal{L} a^3\frac{\partial P}{\partial z}}{3 \mu_e},
\end{align}
hence, from consistency, we have: 
\begin{align}
\label{eq:ApB:19}
\mu_e = \mu\left[ 1 - \frac{6\beta}{\kappa^2a^2\Fcc}f\left(\kappa a,\beta\right) \left(1 - \frac{1 }{\kappa a } \tanh(\kappa a) \right)\right]^{-1}.
\end{align}
Note that $\mu_e$ also have the following functional definition
\begin{align}
\label{eq:ApB:20}
\frac{\mu_e}{\mu} = \frac{Q_{z,0}}{Q_{z,n_{\infty}}}
\end{align}
Where the extra  subscript on $Q_z$ denotes the ion number density. 

\section{First-order amplitude expansion of the flow field in the absence of surface charge}
\label{sec:smallamp}
Here we expand the flow field to the first order in a small surface undulation.
We consider first the case where there is no surface charge.
Hence, we consider a system and solution independent of the $z$ coordinate; considering for simplicity the domain between $y=0$ and $y=h(x)$.
Without loss of generality, the domain has been inverted along $y$ compared to the numerical simulations.
The surface undulation function is given by $h(x) = H(1+\epsilon \cos kx)$.
We seek an expression which is first-order in $\epsilon$ for the flow field ($\v u = u_z \hat{\v z}$).
The equation to solve is the Poisson problem
\begin{equation}
	\laplacian u_z = -f
    \label{eq:uzf}
\end{equation}
with the no-slip condition $u_z=0$ on the top and bottom boundaries specified above.
Following \cite{tsangaris1984}, we make the coordinate transformation
\begin{align}
	\eta  &= x, \\
    \zeta &= \frac{y}{h(x)},
\end{align}
and in these coordinates, the domain is $\zeta \in [0, 1]$.
In the new coordinates, the Laplace operator is given by \cite{tsangaris1984}:
\begin{multline}
	\laplacian = \frac{1}{h^2}\left( 1 + \zeta^2 h_\eta^2 \right) \partial_{\zeta\zeta} - 2 \zeta \frac{h_\eta}{h} \partial_{\zeta\eta} + \partial_{\eta\eta} \\
    + \zeta \left( 2 \left(\frac{h_\eta}{h}\right)^2 - \frac{h_{\eta\eta}}{h}\right) \partial_\zeta,
\end{multline}
so to first order in $\epsilon$, Eq.\ \eqref{eq:uzf} gives (letting $u_z = u^{(0)} + \epsilon u^{(1)}$)
\begin{multline}
	H^2 \laplacian u_z = u_{\zeta\zeta}^{(0)} + \Lambda^2 u_{\beta\beta}^{(0)} 
    	+ \epsilon \Big[ u_{\zeta\zeta}^{(1)} - 2 \cos \beta u_{\zeta\zeta}^{(0)} \\+ 2 \zeta \Lambda^2 \sin \beta + \Lambda^2 u_{\beta\beta}^{(1)} +\Lambda^2 \sin \beta \cos \beta u_\zeta^{(0)} \Big] = -f
\end{multline}
where $\beta = k \eta$, and $\Lambda = k H$ characterizes the ratio between channel height and wavelength of the undulation.
Solving this to zeroth and first order gives the solution
\begin{align}
	u^{(0)} &= \frac{f}{2} \zeta \left( 1 - \zeta \right) \\
    u^{(1)} &= \frac{f}{2} \left[ (1-2\zeta) \zeta + \frac{\sinh\Lambda\zeta}{\sinh\Lambda} \right] \cos \beta
\end{align}
and, hence, the full expression in the original coordinates becomes
\begin{equation}
	u_z = \frac{f}{2} \left[ \frac{y}{H} \left( 1 - \frac{y}{H} \right) + \frac{\sinh \frac{\Lambda y}{H} }{\sinh\Lambda} \epsilon \cos k x \right],
    \label{eq:pert_u_z}
\end{equation}
to the first order in $\epsilon$.

Integrating \eqref{eq:pert_u_z} over $y$ and subsequently over the narrowest region, $x \in [L_x/4, 3L_x/4]$ yields (note that the $x$-axis is shifted compared to the numerical simulations), according to the definition \eqref{eq:def_QzA} of $Q_{z,x}(A,0)$:
\begin{equation}
	Q_{z,x} (A, 0) = \frac{fHL_x}{2} \left[ \frac{1}{12} - \frac{\epsilon}{\pi \Lambda} \left( 1 - \frac{1}{\cosh\Lambda} \right) \right],
\end{equation}
while the total flux is given by:
\begin{equation}
	Q_{z,t} (A, 0) = \frac{fHL_x}{12}.
\end{equation}
This yields, using Eq.~\eqref{eq:def_ThetaA}, the absolute asymmetry:
\begin{equation}
	\Theta_{x}(A,0) = \frac{1}{2} - \frac{6 \epsilon}{\pi \Lambda} \left( 1 - \frac{1}{\cosh\Lambda} \right) .
\end{equation}
Finally, identifying $H=L_y$, $A = L_y\epsilon $ and $\Lambda = 2\pi L_y/L_x$, we may write this in the somewhat more familiar form:
\begin{equation}
	\Theta_{x}(A,0) = \frac{1}{2} - \frac{6 A L_x}{2\pi^2 L_y^2} \left[ 1 - \frac{1}{\cosh\left( 2\pi \frac{L_y}{L_x} \right)} \right] .
\end{equation}

It is interesting to note that corrections to $Q_{z,t} (A,0)$ and $Q_{z,y}(A, 0)$ are both of at least order $\mathcal O (\epsilon^2)$, and hence the curves plotted against $A$  should be flat at $A=0$.

\bibliography{references}

 \end{document}